\documentclass[aps,pra,reprint,showpacs,amsmath,amssymb,groupedaddress]{revtex4-1}

\usepackage{graphicx}
\usepackage{epstopdf}
\usepackage{dcolumn}
\usepackage{bm}
\usepackage{tabularx}
\usepackage{multirow}
\usepackage{lipsum}

\begin{document}

\preprint{}

\title{Core-Shell Magneto-Optical Trap\\ for Alkaline-Earth-Metal-Like Atoms}

\author{Jeongwon Lee, Jae Hoon Lee, Jiho Noh and Jongchul Mun}

\affiliation{Center for Time and Frequency, Korea Research Institute of Standards and Science, Daejeon 305-340, Korea}

%\date{\today}

\begin{abstract}
We propose and demonstrate a new magneto-optical trap (MOT) for alkaline-earth-metal-like (AEML) atoms where the narrow $^{1}S_{0}\rightarrow$$^{3}P_{1}$ transition and the broad $^{1}S_{0}\rightarrow$$^{1}P_{1}$ transition are spatially arranged into a core-shell configuration. Our scheme resolves the main limitations of previously adopted MOT schemes, leading to a significant increase in both the loading rate and the steady state atom number. We apply this scheme to $^{174}$Yb MOT, where we show about a hundred-fold improvement in the loading rate and ten-fold improvement in the steady state atom number compared to reported cases that we know of to date. This technique could be readily extended to other AEML atoms to increase the statistical sensitivity of many different types of precision experiments.
\end{abstract}

\maketitle

Alkaline-earth-metal-like (AEML) atoms with two valence electrons have received great interest in various fields~\cite{Hinkley2013, Bloom2014, Poli2011, Zhang2014, Rosenband2008, Chou2010, Daley2008, Gerbier2010, Natarajan2005, Guest2007} due to its spinless ground state $^{1}S_{0}$, and a long-lived metastable state $^{3}P_{0}$. More specifically, precision spectroscopy of lattice-trapped AEML atoms opened up many possibilities such as optical clocks~\cite{Hinkley2013, Bloom2014}, gravimeters~\cite{Poli2011}, and most recently, the exploration of SU(N) magnetism~\cite{Zhang2014}. Another branch of experiment uses laser cooled AEML atoms to search for the permanent electric dipole moment~\cite{Natarajan2005, Guest2007}. For many of these precision measurement experiments, increasing the atom number in the cold atomic sample and/or reducing the preparation time directly affect the statistical sensitivity. Therefore, a fast-loading magneto-optical trap (MOT) with high atom numbers becomes desirable.

\begin{figure}
\includegraphics[width = 3.3 in]{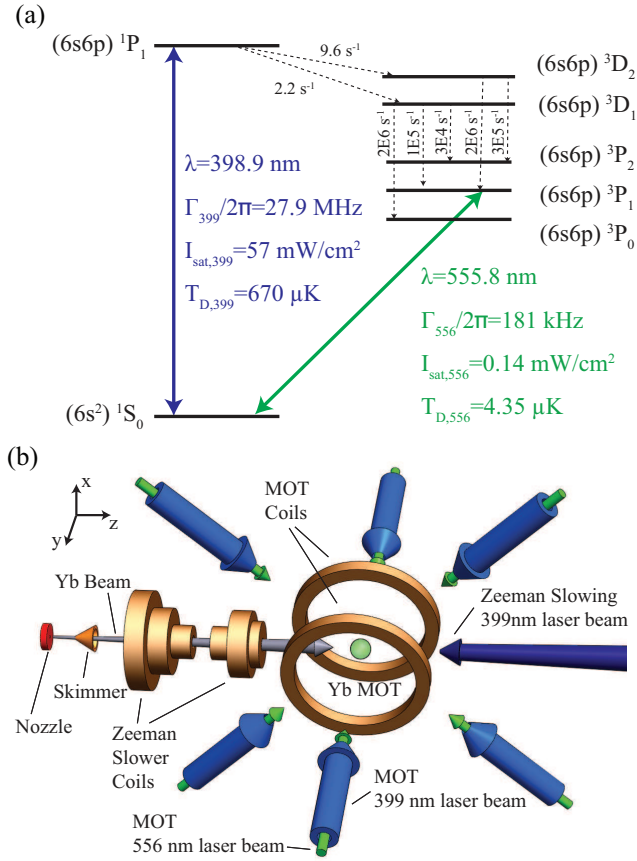}
\caption{(Color online) (a) Energy level diagram of ytterbium showing the singlet and the triplet laser cooling transition along with relevant physical properties. $\Gamma$ is the natural linewidth, $I_{sat}$ is the saturation intensity, and $T_{D}$ is the Doppler temperature of the cooling transition. The dotted line shows the various decay channels of $^{1}P_{1}$ state. (b) Schematic view of the experimental setup for the core-shell magneto-optical trap.}
\label{f:ShellMOTSetup}
\end{figure}

As one of the AEML atoms, Yb has two commonly used laser cooling transitions(Fig.~\ref{f:ShellMOTSetup}(a)), which are the broad $^{1}S_{0}\rightarrow$$^{1}P_{1}$ transition at 399 nm (will be referred as the singlet transition), and the narrow $^{1}S_{0}\rightarrow$$^{3}P_{1}$ transition at 556 nm (will be referred as the triplet transition). Due to the broader natural linewidth of the singlet transition, higher capture velocity and faster loading is allowed for the singlet MOT compared to the triplet MOT. On the other hand, the triplet MOT has advantages over the singlet MOT in terms of lower Doppler temperature limit and perfect cycling of the cooling transition~\cite{Kuwamoto1999}.

There have been two main approaches for efficient loading of Yb atoms into a MOT. One way is to initially load a singlet MOT, and subsequently transfer to a triplet MOT with a different magnetic field gradient~\cite{Maruyama2003, Poli2008, Pandey2010}. While one could achieve fast loading using this transfer technique, the steady state atom number of the singlet MOT is limited by the branching ratio of the $^{1}P_{1}$ state decaying to the metastable triplet states as shown on Fig.~\ref{f:ShellMOTSetup}(a). Therefore, to date, the maximum number of trapped atoms has been limited to the $10^7$ range even with high transfer efficiencies~\cite{Maruyama2003}. More recently, a repumping scheme was proposed to remedy the loss of atoms into the metastable triplet states. However, only a limited gain in the atom number of 30\% was reported~\cite{Cho2012}. A second approach is to artificially frequency broaden the triplet cooling laser light via modulation in order to increase the capture velocity~\cite{Kuwamoto1999, Takasu2003, Hansen2011}. Recently, an Yb MOT loading rate of $\sim6\times10^6$ atoms/s has been reported~\cite{Dorscher2013} using this broadband technique. Although this broadband technique results in slower loading compared to the transfer technique, the absence of leaky channels in the triplet transition cooling process allows for higher trapped atom numbers up to $10^8$ range. We note that many of the AEML atoms such as strongtium~\cite{Katori1999}, calcium~\cite{Kraft2009}, radium~\cite{Guest2007}, and erbium~\cite{Frisch2012} share similar MOT schemes based on the singlet and the triplet cooling transitions, where similar limitations are present.

In this paper, we introduce a core-shell MOT scheme, which produces a fast loading MOT with high steady state atom numbers. The basic idea for our scheme is shown in Fig.~\ref{f:ShellMOTSetup}(b). We use the broadband triplet cooling light at the central core part of the MOT (BB triplet core), and surround it with a shell of the singlet cooling light (singlet shell). Unlike the singlet-triplet MOT transfer scheme, we fix the magnetic field gradient at an optimum value throughout the cooling process. This arrangement of the light fields allows us to take advantage of the different properties of the two cooling transitions. The singlet shell with a relatively higher capture velocity increases the MOT loading rate, complementing the BB triplet core with lower capture velocity. The atoms in the singlet shell are then pushed into the BB triplet core where they are further cooled without any limitations caused from the branching ratio mentioned above.

One of the main challenges of this new scheme is to operate both the singlet shell and the BB triplet core under the same magnetic field gradient condition. To first order, we could define an optimum magnetic field gradient as when the Zeeman shift becomes equal to the natural linewidth of the cooling transition at the boundary of the MOT~\cite{Steane1992}. Therefore, with a large linewidth difference between the singlet and the triplet transitions, optimum magnetic field gradients for each transition would be different by more than an order of magnitude. Understanding this problem, we numerically simulate the atomic capture process of our scheme to search for a set of magnetic field gradient and laser frequency detuning parameters which maximizes the loading rate. With the appropriate parameters found from the simulation, we were able to load $\gtrsim1\times 10^9$ Yb atoms in the trap with a loading time constant of $1.2$ seconds.

Our experimental setup is shown in Fig.~\ref{f:ShellMOTSetup}(b), which consists of an oven, skimmer, spin-flip Zeeman slower, and the core-shell MOT. We begin the experiment by heating Yb chunks up to 400 $^{\circ}$C inside the oven. The atoms come out as an effusive beam with a mean axial velocity of around 300 m/s after passing through a capillary type nozzle, which is 10 mm long and 1 mm in diameter. The atomic beam further goes through a motorized atomic beam shutter, a conical shape skimmer with a 2 mm diameter hole, and a differential pumping stage before it enters a 30 cm-long spin-flip Zeeman slower. The total distance between the oven and the MOT is about 90 cm. Under the beam operation, the vacuum pressures were maintained at $\sim10^{-8}$ Torr between the nozzle and the skimmer, and $\sim10^{-11}$ Torr inside the MOT chamber.

The spin-flip Zeeman slower was designed to have a capture velocity of 300 m/s, operated with the singlet transition. An external cavity diode laser was combined with a tapered amplifier to generate 798 nm light, which we frequency doubled with a second harmonic generation unit(Toptica SHG Pro) to get 399 nm light. We use 60 mW of the 399 nm laser beam at a frequency detuning of -640 MHz(-22 $\Gamma_{399}$) for the Zeeman slower. The Zeeman slower laser beam counter-propagates with respect to the atomic beam as shown in Fig.~\ref{f:ShellMOTSetup}(b), and is focused down from 10 mm at the MOT center to 1 mm diameter at the oven nozzle in order to match the atomic beam divergence. The final axial velocity distribution of our Zeeman slower was measured to be centered around 25 m/s using laser induced fluorescence spectroscopy. In general, the final atomic beam velocity after the Zeeman slower should be less than the MOT capture velocity for an efficient MOT loading.

\begin{figure}
\includegraphics[width = 3.3 in]{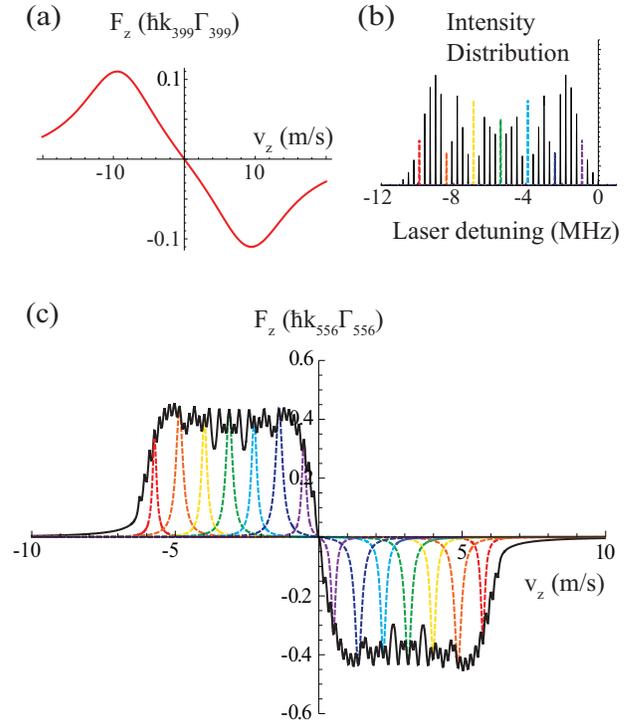}
\caption{(Color online) (a) Calculated velocity dependent radiation force for the singlet transition with $s=0.32$, $\delta_{L,399}=-45 MHz$, and $dB/dz=5 G/cm$ shown in red solid line. (b) and (c) Calculated frequency dependent light intensity distribution and velocity dependent radiation force for BB triplet transition under $s=400$, $dB/dz=5$ G/cm, and the laser frequency modulation conditions described in the text. The dashed lines in different colors show the multiple radiation force terms coming from 7 sampled frequency components of the frequency modulated triplet transition. Black solid line shows the total radiation force coming from all frequency components.}
\label{f:ForceCurves}
\end{figure}

After the Zeeman slower, the atoms are captured by the core-shell MOT where we have two spatially separated coaxial laser beams counter propagating along all 3 dimensions as described earlier and depicted in Fig.~\ref{f:ShellMOTSetup}(b). For each axis, we use a circular mask with a chosen diameter to block the center of a 10 mm $1/e^{2}$-diameter singlet transition laser beam, while the hole is filled in with a matching $1/e^{2}$-diameter triplet transition laser beam using a dichroic mirror. Each combined beam propagates for up to 80 cm including the retro-reflected distance to create the MOT. We use a 399 nm diode laser for the singlet shell. The singlet transition laser beam powers after the mask were 15 mW for each of the two axes on the \textit{x}-\textit{z} plane (Fig.~\ref{f:ShellMOTSetup}(b)), and 3 mW along the \textit{y}-axis. As for the BB triplet core, a frequency doubled fiber laser(Menlo Orange One and PPLN SHG unit) light at 556 nm was passed through a sinusoidally frequency modulated acousto-optic modulator for the frequency broadening effect. The power is divided to deliver 20 mW for each of the two axes on the \textit{x}-\textit{z} plane, and 4 mW along the \textit{y}-axis. Laser frequency detunings and the magnetic field gradient value are chosen from the numerical results shown below.

We start by analyzing the radiation force experienced by the atoms at its steady state, for two different cooling transitions. As for the singlet cooling transition, we write down the radiation force in terms of the saturation parameter $s=I/I_{sat}$ and the total detuning $\Delta$,

\begin{equation}\label{eq:1}
\vec{F}(s,\Delta )=\rho _{ee}(s,\Delta ) \times \Gamma \hbar \vec{k},
\end{equation}
where $\rho _{ee}$ is the steady state excited state population, $\Gamma$ is the natural linewidth of the transition, and $\vec{k}$ is the wavevector of the light field. Considering only the z-direction, the total detuning parameter can be broken down to, $\Delta=\delta_{L}-k_{z}v_{z}-\mu _{B}\partial B/\partial z \times  z/\hbar $, where $\delta_{L}$ is the laser detuning, $k_{z}v_{z}$ is the Doppler shift from the moving atoms, and $\mu _{B}\partial B/\partial z \times  z/\hbar$ is the position dependent Zeeman shift. In Fig.~\ref{f:ForceCurves}(a), we use Eq.~(\ref{eq:1}) to plot the velocity dependent radiation force coming from the singlet transition laser beam exerted along the z-direction, with typical experimental conditions of $s=0.32$, $\delta_{L,399}=-45 MHz$, and $dB/dz=5 G/cm$.

In order to calculate the radiation force of the broadband triplet cooling transition, we consider a 2-level atom interacting with multiple light fields at different detunings. In case of sinusoidal frequency modulation of the laser, we can control the modulation frequency $f_{m}$, and the modulation index $H$. Fig.~\ref{f:ForceCurves}(b) shows the Bessel-type intensity distribution of the frequency modulated spectrum for the conditions of $\delta_{L,556}$=-5.4 MHz, $f_{m}$=300 kHz, and $H$=13. We first show how individual frequency components of the modulated spectrum exert force on atoms in different velocity classes. In Fig.~\ref{f:ForceCurves}(c), we show multiple velocity dependent force curves in colored dash lines corresponding to the 7 different sampled frequency components shown with the same color in Fig.~\ref{f:ForceCurves}(b). Here we used conditions of $s=400$, $dB/dz=5$ G/cm, and the aformentioned frequency modulation parameters.  Next we calculate the atomic interaction when all of these frequency components are present at the same time. The problem has been studied by solving the optical Bloch equations after separating out the Fourier frequency components~\cite{Savels2005}. Following this approach we write the steady state excited state population of the broadband triplet transition, $\rho _{ee,BB}$, in terms of the multiple frequency components as,

\begin{equation}\label{eq:2}
\rho _{ee,BB}=\frac{\int_{0}^{\infty }d\omega J[\omega ]\frac{1}{\Delta ^{2}+(\Gamma /2)^{2}}}{1+2\int_{0}^{\infty }d\omega J[\omega ]\frac{1}{\Delta ^{2}+(\Gamma /2)^{2}}},
\end{equation}
where $J[\omega]$ is the spectral density function. Furthermore, $J[\omega]$ can be defined in terms of the Fourier transform of the time dependent Rabi frequency, $\Omega (\omega )$, in the following,
\begin{equation}\label{eq:3}
<\Omega (\omega )\Omega (\omega' )>=J[\omega ]\delta (\omega+\omega').
\end{equation}
Combining Eq.~(\ref{eq:1})$\sim$~(\ref{eq:3}) along with the generalized rotating wave approximation~\cite{Savels2005}, we can plot the result of the effective radiation force curve when all the frequency components are present, as shown in Fig.~\ref{f:ForceCurves}(c) with a black solid line.

Comparing Fig.~\ref{f:ForceCurves}(a) and ~\ref{f:ForceCurves}(c), we notice the difference in both the magnitude and the overall shape of the radiation force curves. The peak magnitude of the radiation force from the singlet transition is $\sim$60 times larger than the broadband triplet transition, since there is a large difference between $\Gamma_{399}$ and $\Gamma_{556}$. Due to the spatial arrangement of the light fields in our core-shell MOT scheme, the higher velocity atoms near the edge of the MOT are affected only by the strong singlet transition radiation force. On the other hand, the cooling properties of the lower velocity atoms near the center of the MOT are completely determined by the broadband triplet transition. By varying the frequency modulation parameters of the BB triplet core, we can change the shape of the radiation force curve shown in Fig.~\ref{f:ForceCurves}(c), which consequently affects both the spatial distribution and the temperature of the trapped atoms.

\begin{figure}
\includegraphics[width = 3.3 in]{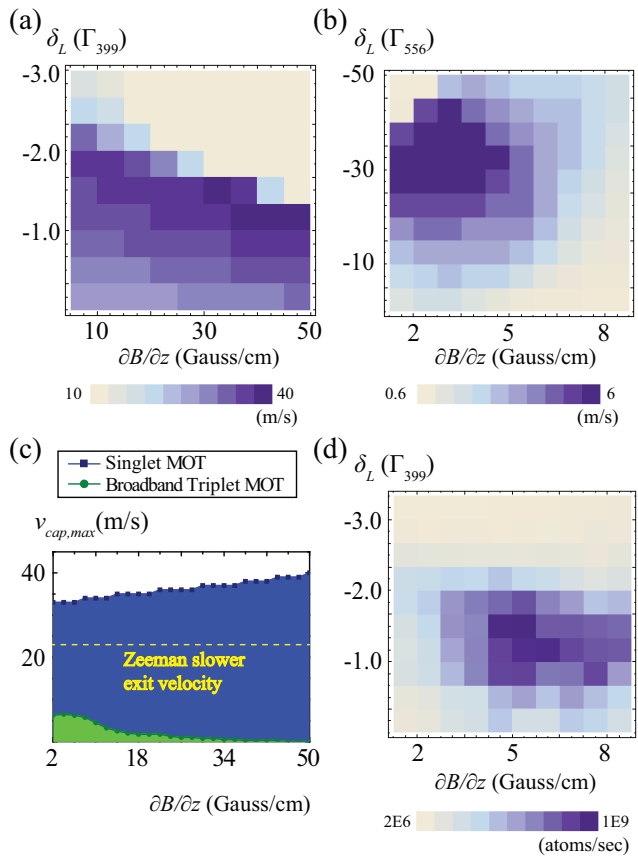}
\caption{(Color online) Numerically simulated capture velocities under various $\partial B/\partial z$ and $\delta_{L}$ for (a) the singlet transition, and (b) the broadband triplet transition, where we kept the frequency modulation parameters to $f_{m}$=300 kHz, and $H$=13. (c) Numerically simulated maximum capture velocity, $v_{cap,max}$, which is the largest capture velocity obtainable at a fixed $\partial B/\partial z$, for both the singlet and the broadband triplet transition. Dashed yellow line shows the Zeeman slowed atomic beam velocity. (d) Experimentally measured core-shell MOT loading rates under the conditions described in the text.}
\label{f:LoadingOptimization}
\end{figure}

Given the calculated radiation force exerting on an atom with a specific position and velocity, we could numerically simulate the atomic movement inside the trapping region using the classical equation of motion. For an atom at displacement $\vec{d}_n=(x_{n},y_{n},z_{n})$ with a velocity $\vec{v}_n=(v_{x,n},v_{y,n},v_{z,n})$ at the $n^{th}$ step of the iteration there is a well-defined radiation force being applied, which can be used to calculate $\vec{d}_{n+1}$ and $\vec{v}_{n+1}$. We take the initial displacement as $\vec{d}_0=(0,0,-R_{MOT})$, where $R_{MOT}$ is the radius of the MOT beam, and the initial velocity as $\vec{v}_0=(0,0,v_{slower})$, which is the velocity of the atom entering the MOT. We iteratively apply the radiation force in unit time steps set to the inverse of the spontaneous emission rate of the transition. We continue the iteration and track both the position and the velocity information of the atom until the simulation is terminated, when either the magnitude of the velocity $\vec{v}_n$ is less than the Doppler limited velocity of the transition(implying successful capture), or the displacement $\vec{d}_n$ steps outside the spherical boundary of the laser beam(implying failure of capture). With this numerical simulation tool, we could retrieve the largest initial velocity $v_0^{*}$ of atoms that are captured. We define this as the effective capture velocity, $v_{cap}$, for the corresponding sets of simulation conditions.

We plot the simulated capture velocities for the singlet MOT without the hole at the center in Fig.~\ref{f:LoadingOptimization}(a), and the results for broadband triplet MOT in Fig.~\ref{f:LoadingOptimization}(b). We used two control parameters in each plot, which are the magnetic field gradient and the laser frequency detuning for the MOT, while fixing all the other parameters as it was described in Fig.~\ref{f:ForceCurves}. The highest capture velocity of the singlet MOT is 40 m/s with $\partial B/\partial z=50$ G/cm, while the BB triplet MOT capture velocity peaks at 5 m/s with $\partial B/\partial z=5$ G/cm. The difference in the maximum capture velocities and the optimum $\partial B/\partial z$ values can be understood by the aforementioned natural linewidth dependence.

In Fig.~\ref{f:LoadingOptimization}(c), we plot the largest capture velocity obtainable for various cooling laser detunings at a fixed $\partial B/\partial z$, $v_{cap,max}$, for both the singlet and the BB triplet MOT. From the plot, we observe different behaviors for the two types of MOTs in terms of how the capture velocity decreases as we move away from the optimum value of $\partial B/\partial z$. As for the singlet MOT, we see that the maximum capture velocity is always above our Zeeman slowed atomic beam velocity, within the $\partial B/\partial z$ values studied here. On the other hand, the maximum capture velocity of the BB triplet MOT drops more than 2-fold when $\partial B/\partial z>10$ G/cm. These results indicate that we should operate the core-shell MOT closer to the optimum magnetic field gradient of the BB triplet MOT, since the singlet MOT is less sensitive to the magnetic field gradient in terms of its capture velocity.

\begin{figure}
\includegraphics[width = 3.3 in]{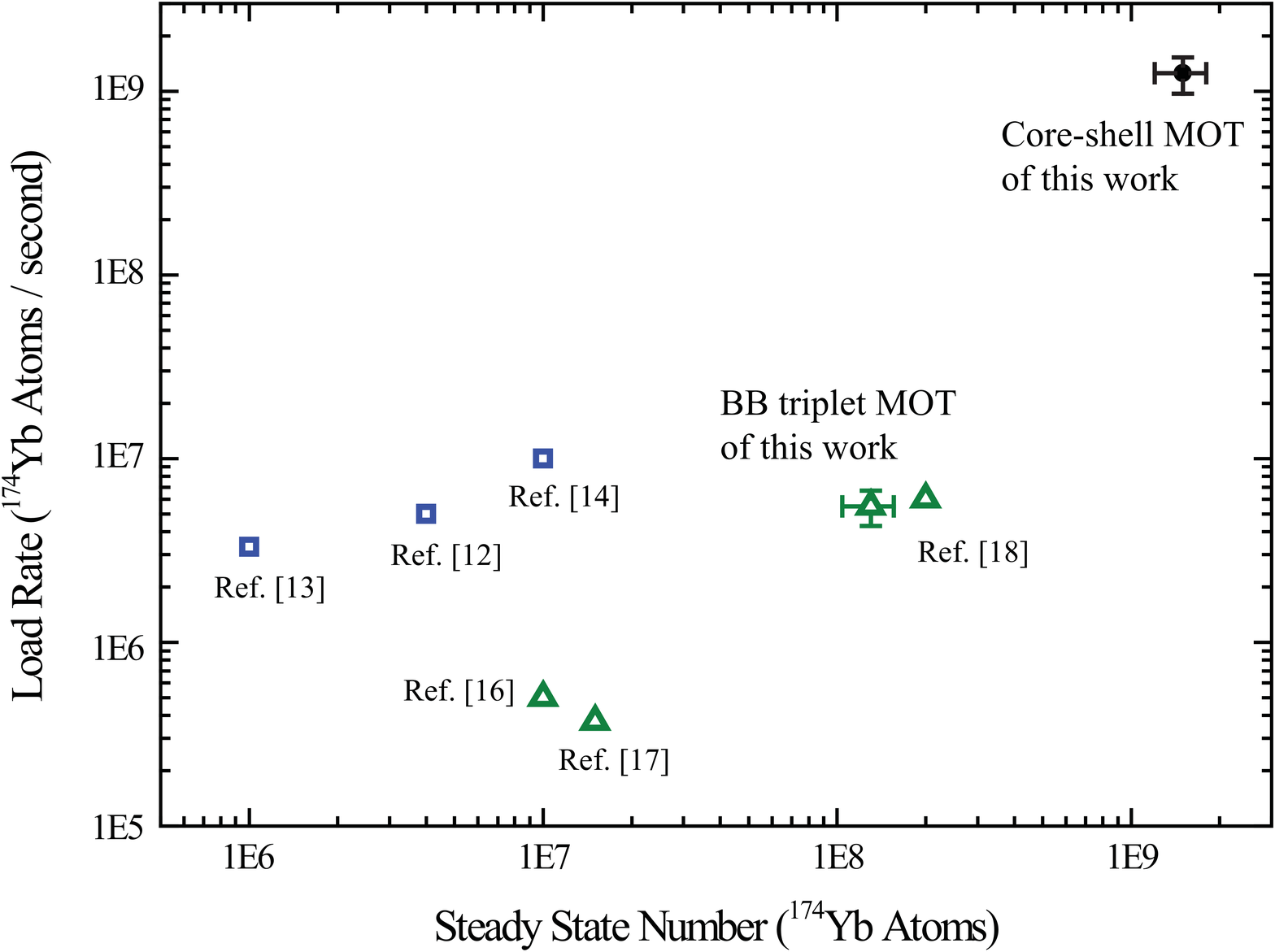}
\caption{(Color online) Loading rate and steady state atom number comparison between the core-shell MOT and the previous results. Blue open squares represent previous experiments using the singlet-triplet MOT transfer scheme. Green open triangles show the experiments using the broadband triplet MOT scheme, including experimental results for our setup.}\label{f:LoadRateComparison}
\end{figure}

With the information obtained from Fig.~\ref{f:LoadingOptimization}(c), we could narrow down the experimental search range for the laser frequency detunings and $\partial B/\partial z$ values, which would give the highest loading rate for the core-shell MOT scheme. In Fig.~\ref{f:LoadingOptimization}(d), we show the measured core-shell MOT loading rates under various experimental conditions. We fixed the laser parameters of the BB triplet core at $s=400$, $f_{m}$=300 kHz, $H$=13, $\delta_{L,556}$=-5.4 MHz(-30 $\Gamma_{556}$), while varying the magnetic field gradient and the laser frequency detuning of the singlet shell. The loading rate is described by the rate equation~\cite{Dinneen1999},

\begin{equation}\label{eq:4}
\dot{N}=L-\alpha N-\beta \int n^{2}(\mathbf{r})d^{3}\mathbf{r}
\end{equation}
where $L$ is the loading rate, $\alpha$ is the linear loss rate, $\beta$ is the two body collision coefficient, $n(\mathbf{r})=n_{0}e^{-(\mathbf{r}/a)^{2}}$ is atomic density distribution in the trap with $n_{0}$ as the peak density, and $a$ represents the width of the atomic cloud. Solving the rate equation with a finite loading rate gives,

\begin{equation}\label{eq:5}
N(t)=N_{ss}\frac{1-e^{-\gamma t}}{1+\xi e^{-\gamma t}},
\end{equation}
where $N_{ss}$ is the steady state number, $\gamma =\alpha +\beta n_{0}/\sqrt{2}$ is the total trap loss rate, and $\xi =\beta n_{0}/(\beta n_{0}+2\sqrt{2}\alpha )$ is the collision loss fraction. We extract $\gamma$ and $\xi$ from the time dependent fluorescence data recorded by a photo-multiplying tube (PMT). $N_{ss}$ and the density distribution of the atoms within the MOT were measured from an absorption image of a singlet transition probe beam, where we have used a charged-coupled device (CCD) camera. Combining the measured values, the experimental loading rates are given by $L=\alpha N_{ss}+\beta (\sqrt{2\pi }a)^{-3}N_{ss}^{2}$. The highest loading rate of the core-shell MOT was achieved when $\delta_{399}=$-1.6 $\Gamma_{399}$ for the singlet shell, $\delta_{556}=$-30 $\Gamma_{556}$ for the BB triplet core, and $\partial B/\partial z=5$ G/cm for the magnetic field gradient as shown in Fig.~\ref{f:LoadingOptimization}(d). At the optimal setting, we have trapped up to $N_{ss}=1.5(3)\times10^9$ $^{174}$Yb atoms with a loading time constant of $N_{ss}/L=1.2(1)$ seconds. As shown in Fig.~\ref{f:LoadRateComparison}, our result shows more than two orders of magnitude improvement in terms of the the loading rate and ten-fold improvement in terms of the steady state atom number compared to most reported cases of $^{174}$Yb MOTs. After the core-shell MOT loading process reaches its steady state, we follow Ref.~\cite{Kuwamoto1999} to reduce the laser-broadening amplitude of the BB triplet core and achieve final MOT temperature of $\sim20\mu K$, which is comparable to other Yb experiments~\cite{Kuwamoto1999, Dorscher2013, Hansen2011}.

\begin{figure}
\includegraphics[width = 3.3 in]{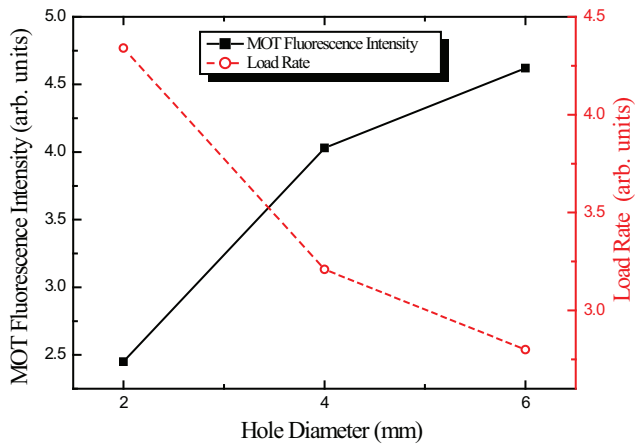}
\caption{(Color online) Relative change in steady state MOT fluorescence intensity and loading rates for the core-shell MOT with different hole diameters.}
\label{f:HoleDiameterChange}
\end{figure}

We also study how the thickness of the singlet shell within our MOT geometry affects the loading rate and the steady state atom number. For this purpose, we varied the circular mask diameter which creates different size holes at the center of the singlet transition MOT beams. The triplet transition MOT beam diameters were changed accordingly to fill the hole, while the intensity was kept constant. Fig.~\ref{f:HoleDiameterChange} shows the relative change in the steady state MOT fluorescence intensity and the loading rate with 3 different hole sizes. We observed faster loading rates as we decrease the hole diameter, which is consistent with our intuition since it creates a thicker outer shell of the strong singlet transition having more cooling power. On the other hand, the steady state atom number in the MOT decreases with smaller central core volume, which is represented by the decrease in steady state MOT fluorescence intensity. Such behavior is due to the increase in atom loss rates as the density dependent term in Eq.~(\ref{eq:4}) becomes larger with smaller volumes.

\begin{figure}
\includegraphics[width = 3.3 in]{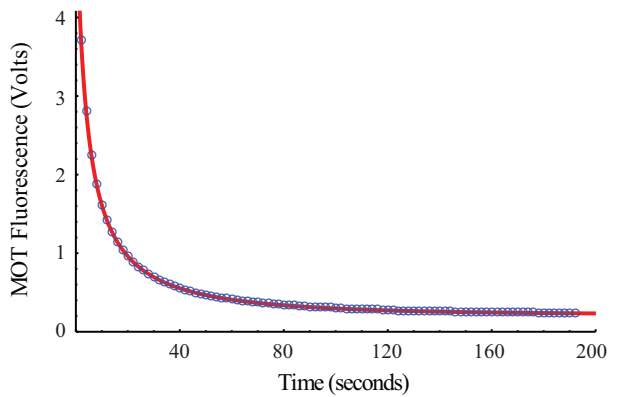}
\caption{(Color online) Experimental measure of MOT fluorescence decay rate. The decay fluorescence are measured at 100,000 samplings/second and binned with 1 second intervals, shown as the blue open circles. The red curve is the density dependent decay fit giving a linear loss rate of $\alpha=0.017(1)s^{-1}$ and a two body loss coefficient of $\beta=4.2(5)(16)\times 10^{-12}cm^{3}s^{-1}$.}
\label{f:DensityDecay}
\end{figure}

Utilizing the fast loading of the core-shell MOT scheme, we were able to reach densities of $\gtrsim1\times10^{11}$ $^{174}$Yb atoms/$cm^3$ in the trap, which allowed us to observe the two body collisional effects from the trap loss behavior. In Fig.~\ref{f:DensityDecay}, we show the MOT fluorescence decay signal after turning off the atomic beam, the singlet shell beams, and the modulation for the BB triplet core beams. The singlet shell beams were turned off to ensure that the density dependent decay mechanism solely comes from the triplet transition. The triplet transition MOT laser beams were kept at a frequency detuning of -5.4 MHz(-30 $\Gamma_{556}$), and the intensities were 400 $I_{sat,556}$ for each of the two axes on the \textit{x}-\textit{z} plane, and 80 $I_{sat,556}$ along \textit{y}-axis throughout the measurement. The decay in fluorescence are monitored with a photo-multiplying tube at 100,000 samplings/second and binned with 1 second intervals. The MOT is initially in the high density limit where the two body loss term in Eq.~(\ref{eq:4}) is larger than the linear term. Hence, the fast initial decay process is dominated by the $\beta$ term. As the atomic density decreases over time, the linear loss term starts to win over the two body term, where the decay behavior is characterized by the $\alpha$ term. Similar to the way we analyzed the MOT loading, we solve Eq.~(\ref{eq:4}) in the absence of the loading term to get,

\begin{equation}\label{eq:6}
N(t)=\frac{N_{ss}}{\left ( 1+\beta n_{0}/\alpha  \right )e^{\alpha t}-\beta n_{0}/\alpha},
\end{equation}
which can be used to fit the decaying MOT fluorescence. The resulting fit is shown as a red solid line in Fig.~\ref{f:DensityDecay}, which matches well with the experimental data points. Combining the fit coefficients of Eq.~(\ref{eq:6}) and the peak density measured from the absorption profile, we get a linear loss rate of $\alpha=0.017(1) s^{-1}$ and a two body loss coefficient of $\beta=4.2(5)(16)\times 10^{-12} cm^{3}s^{-1}$, where 2$\sigma$ fitting errors are shown inside the first parentheses and the systematic error coming from the absolute density calibration is shown in the second parenthesis. The measured $\alpha$ is believed to be dominated by the background gas collision rate. The two body term $\beta$ is closely linked to the photoassociation (PA) process. The PA process of Yb has been investigated experimentally~\cite{Takasu2004, Enomoto2008} at different density and atom number regimes using optical dipole traps. We believe our core-shell MOT scheme could be used to study the density dependent trap loss rates complimenting the previous results, which could provide more insights to the PA process of AEML atoms.

In this work, we have developed a core-shell MOT scheme for Yb atoms where we utilize both the singlet and the triplet cooling transitions in two spatially separated zones to achieve not only a fast loading rate but also a high steady state atom number in the trap. We studied how the radiation force is exerted for this new type of MOT, and discussed how to set the relevant parameters in order to maximize the loading efficiency. Our scheme was applied to $^{174}$Yb atoms which revealed a hundred-fold faster loading rate and a ten-fold higher steady state atom numbers compared to other reported cases to date. We measured the density dependent decay behaviour of the $^{174}$Yb MOT and extracted the two body loss coefficient, which is relevant to the PA process. In addition, we believe our scheme could be realized in other AEML atoms possessing the singlet and triplet transitions to gain similar benefits. Further investigation of the core-shell MOT scheme in other AEML atoms has potentials to reduce/eliminate the repumping during the MOT stage. The increase in loading rate and trapped atomic number would lead to enhancement of the statistical sensitivity in many different types of precision experiments using cold AEML atoms, such as lattice clock experiments and the electric dipole moment experiments.

\begin{acknowledgments}

We thank C.Y. Park for helpful discussions. This work was supported by KRISS creative research initiative.

\end{acknowledgments}

\appendix*

\bibliography{JLeeCoreShellMOT}

\end{document}